\documentstyle[aps,epsf,prl,twocolumn,multicol]{revtex}
\begin{document}
\draft
\onecolumn
\title{Devil's staircase in kinetically limited growth}

\author{G.J. Ackland}
\address{Department of Physics and Astronomy, Rutgers University, 
Piscataway, New Jersey 08855-0849\\
Permanent address:
Department of Physics and Astronomy, The University of
Edinburgh\\
Edinburgh, EH9 3JZ, Scotland, United Kingdom \\ gjackland @ ed.ac.uk} 
\date{\today}
\maketitle

The devil's staircase is a term used to describe surface or an
equilibrium phase diagram in which various ordered facets or phases
are infinitely closely packed as a function of some model parameter.
A classic example is a 1-D Ising model\cite{bak} wherein long-range
and short range forces compete, and the periodicity of the gaps
between minority species covers all rational values.  In many physical
cases, crystal growth proceeds by adding surface layers which have the
lowest energy, but are then frozen in place.  The emerging layered
structure is not the thermodynamic ground state, but is uniquely
defined by the growth kinetics.  It is shown that for such a system,
the grown structure tends to the equilibrium ground state via a
devil's staircase traversing an infinity of intermediate phases.
It would be extremely difficult to deduce the simple growth law based
on measurement made on such an grown structure.



\begin{multicols}{2}

The original devil's staircase is a footpath between Kingshouse to
Kinlochleven in Scotland, so called because of the huge number of
discrete steps between Glencoe and the ridge.  In technical usage, the
term has been used to describe situations in which the number of
discrete steps within a finite range becomes formally infinite.
Examples include {\it inter alia} the formation of facets of a
crystal\cite{landau,pieranski}, antiferroelectric, smectic and
lyotropic liquid crystals\cite{wang,bahr,pieranski}, magnetic
structure in cerium monopnictides\cite{shibata} and granular
media\cite{combe} Usually the staircase emerges from the interplay
between long-range repulsive (antiferromagnetic) and short range
attractive (ferromagnetic) forces, with transitions between stable
phases appearing as the relative strengths of the interactions are
altered.  The precise form of the interactions is not
crucial\cite{nonconvex}.

The drive toward nanofabrication has led to a tremendous interest in 
growing multilayer structures. In typical methods such as molecular beam 
epitaxy or chemical vapor deposition careful control of the composition 
of the deposited material is required to create complex artificial 
structures.  Without such careful control non-periodic structures tend to 
form.   By contrast, some structures of technological interest 
such as quantum dots may self-assemble, and understanding the 
local equilibria which govern growth is crucial.  In this letter
layer-by-layer surface growth for a simple model is shown to yield
a devil's staircase structure. This suggests that 
for a wide class of systems the expected structure  grown at 
zero temperature is aperiodic, and might easily 
be misinterpreted as disordered.  This apparent disorder is not real, 
but arises from the locally stable structure being dependent on the thickness 
of the film, and there being an infinite number of locally stable structures.

Specifically, the Hamiltonian for our model is equivalent that 
considered by Bak and Bruinsma\cite{bak,bruinsma}:

\begin{equation} 
H = A\sum_i \sigma_i + \sum_{ij} r_{ij}^{-\nu} \sigma_i\sigma_{j}
\end{equation}

This model describes a situation in which each layer can be one of 
two types $\sigma_i=\pm 1$.  The first term gives  $\sigma_i=-1$
a lower formation energy than  $\sigma_i=+1$, while the second term 
gives a long ranged repulsion between like-layers. 

Previous work has concentrated on the devil's staircase as 
an equilibrium phenomenon, and searched for the thermodynamic 
ground state.  Here, by contrast, the dynamics of growth are 
considered, spins being added to the system so as to minimise 
the energy, but then being fixed forever as further layers grow. 

Many physical systems can be mapped onto this Hamiltonian:  a simple example 
is a line of charges in an external field.
The same Hamiltonian describes a situation where the $\sigma_i$
represent the separations between layers rather than the layers
themselves.  Now the first term indicates that it costs less energy to
grow either type on a similar layer, while the second term again
indicates long-range repulsion(attraction) between like(unlike)
layers.  This might describe a system where epitaxial growth was
favored, but generates a long range strain field which needs to be
periodically relieved.  

Alternately, it may describe a situation such as silicon carbide 
growth\cite{rutter1,rutter2} or stacking of close-packed planes, where each
layer is locally either ABA or ABC stacked depending on its neighbors.
Now $\sigma$ represents the relative orientation of adjacent layers.  
In close packed layer (AB) and interlayer ($\sigma$) notation
equivalent  stacking sequences for fcc containing a growth fault are
\begin{eqnarray}
 {\bf ...A\hspace{0.6mm} B\hspace{0.6mm} C \hspace{0.6mm}A\hspace{0.6mm} B \hspace{0.6mm}C }&  {\bf A} & {\bf C\hspace{0.6mm} B\hspace{0.6mm} A\hspace{0.6mm} C\hspace{0.6mm} B\hspace{0.6mm} A...} \\
 ...++++++ &-& ++++++... 
\end{eqnarray}
Notice that a single fault of this type cannot be 
accommodated within periodic boundary conditions.  This led 
Bak and Bruinsma to postulate that the actual defects in the devil's 
staircase are fractional, since more than one must be created together.  
In the growth case there is no such constraint: 
this is equivalent to the difference between intrinsic and extrinsic 
stacking faults in close packed materials (which can arise from removal 
or insertion of a plane) and growth defects (basal plane twins) 
which reverse the sense of stacking and can be generated only by
finite shear of the entire sample or during growth.

Finally, the model can describe a simple history-dependent system,
where the state of the system depends on a sum over its historical
values.  In this case the ``layer number'' should be interpreted as a
time rather than space dimension.

For the growth dynamics, one simply considers adding the 
{\it n+1}th layer to the preexisting {\it n} layers with whichever 
spin reduces the energy.  This can be determined entirely by the 
sign of the local potential.

\begin{equation}
\Delta E_{n+1} = V_{n+1}\sigma_{n+1} =  
(A + \sum_{j=1}^N r_{ij}^{-\nu} \sigma_{j} ) \sigma_i 
\label{eq:pot}
\end{equation}

In zero temperature case considered here, 
if $V_{n+1}$ is positive, the next added layer $ \sigma_{n+1} =  -1$, 
otherwise $ \sigma_{n+1} = +1$.  furthermore, the final 
structure is uniquely defined and while it may appear random, it has zero entropy.

At thermodynamic equilibrium, or asymptotically for the growth dynamics, 
this model (equation 1) has two simple limiting cases. 
For weak long-range interactions 
defined by 

\begin{equation} A  >  \left [ \sum_{n=1}^\infty  n^{-\nu} \right] =
 \zeta(\nu) \end{equation}

  $ \sigma_{n} =  -1$ for all $n$,  meanwhile for small
$A$ alternating behavior  $ \sigma_{n} =  (-1)^n$ is observed.  For 
intermediate values of $A$, the devil's staircase of phases is recovered 
in the asymptotic limit (figure 1)\cite{bak}.

\begin{figure}[ht]
\leavevmode 
\epsfxsize=85mm
\begin{center}
\epsffile{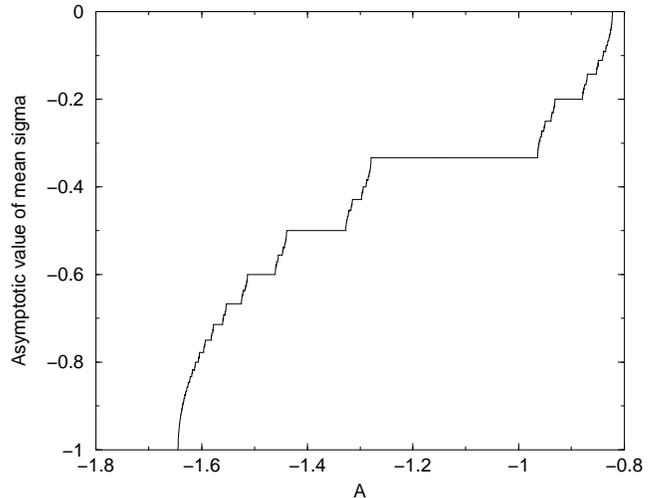}
\end{center}
\caption{Self-similar devil's staircase of phases reached asymptotically 
in growth 
with $\nu=2$.  Plotted are the value of $A$ 
and the mean values of $\sigma$  evaluated over the final 2520 layers 
of a 300000 layer sample.
2520 $( = 2 \times 3 \times 2 \times 5 \times 7 \times 2 \times 3)$
is chosen because it is commensurate with all periodicities 
from one to ten layers, and with 12,14,15,18,20,21 etc).  For these case  
the plotted value of $<\sigma>$ is exact, for others it may be out by
up to 0.04\%. 
\protect\label{fig:asymp}}
\end{figure}

In the case of growth, we find that a convenient parameter to monitor 
is $<V>_m$, the rolling mean value of the potential between $V_{n-m}$ and 
$V_{n}$ wherever a layer type $\sigma_n=+1$ is grown.  
For the case of $\nu=2$ the asymptotic value of  $<V>_{2520}$ plotted against 
$A$ picks out the conventional devil's staircase behavior (Figure 1).

\begin{figure}[ht]
\leavevmode \epsfxsize=80mm
\begin{center}
\epsffile{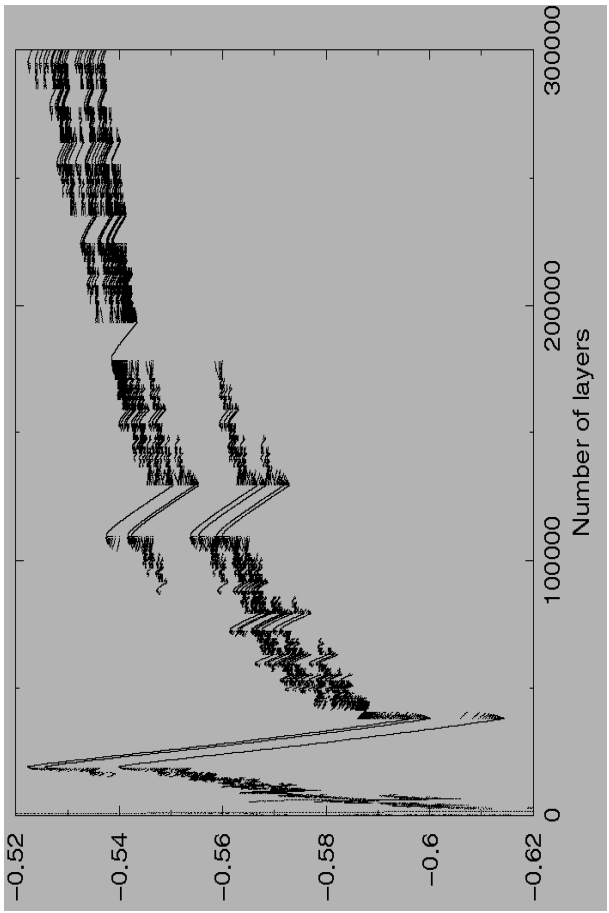}
\end{center}
\caption{Mean surface potential averaged over the preceding 72 layers
before growing the $i^{th}$ $\sigma=+1$ layer.  Highly ordered regions 
correspond to short-period phases which are stable over a significant range of thickness.  The slope of the graph shows that even within these regions, equilibrium has not been reached and ultimately the order breaks down as a new phase is stabilized.  A long range trend towards
an asymptotic value of $<V>_{72}$ is observed.  This figure generated for 
$\nu=1$, A=1, V(0)=0
\protect\label{fig:pot}}
\end{figure}

Our interest lies in the convergence of the structure with layer  number 
- physically how thick a film must grow to recover bulk behavior.  
Again this can be monitored using  $<V>_m$, now plotted against layer number.
For $\nu=2$ the growth converges fairly rapidly to the equilibrium value,
 the effective screening of the surface is fast compared 
with the integer layer spacing.  For smaller $\nu$ convergence is 
slow: Figure \ref{fig:pot} shows the case of  $\nu = 1, A=1$: 
now the  screening is 
sufficiently slow that a wide  range of different phases from the 
``devil's staircase'' are actually observed over a number 
of layers.

\end{multicols}
\begin{figure}[ht]
\leavevmode \epsfxsize=180mm
\begin{center}
\epsffile{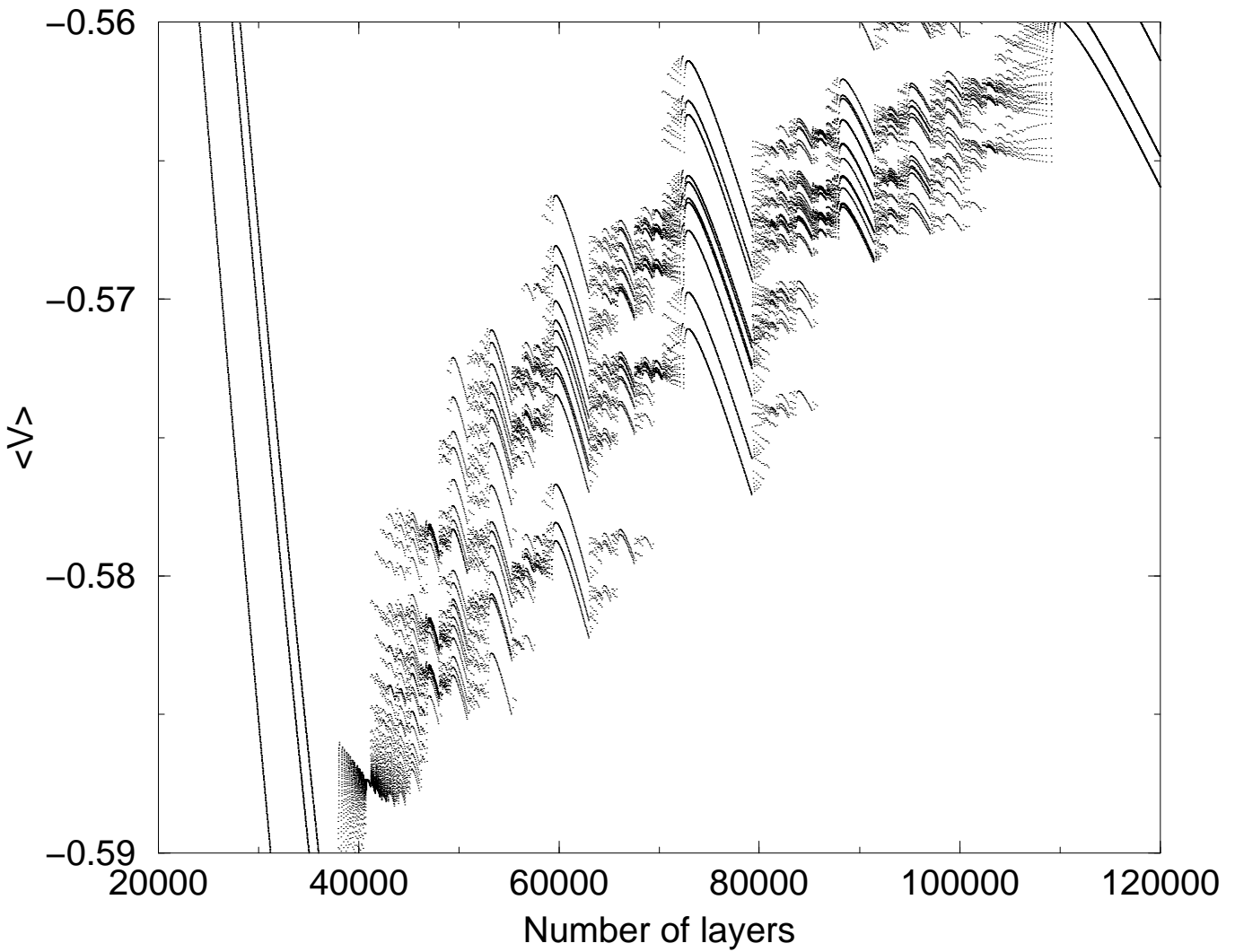}
\end{center}
\caption{Detail from figure 2, illustrating the ordered nature of
of the potential variation for a region where long-period phases 
are stable.  The long-period repeats give rise to multiple values of 
$<V>_{72}$ within the same phase, hence the multiple branches.  This 
multiplicity can be reduced or eliminated by considering $<V>_{2520}$ 
at the expense of smearing out details.  
In this respect the figure is not self-similar.  The (arbitrarily-chosen) 
72-layer averaging is significant on the scale of this figure, 
manifesting itself for the phases which 
are stable over more than 72 layers as an initial increase and 
subsequent curvature in $<V>_{72}$. 
The pure phase behavior is typically a linear
decrease of $<V>_{72}$ with $N$, as seen after 72 layers 
for those  phases which are stable over a sufficiently long period.
\protect\label{fig:pot1}}
\end{figure}

\begin{multicols}{2}

The actual phases and ``screening'' effect are illustrated by the 
running average of the ratio of $\sigma=+1$ to $\sigma=-1$ over the 
previous 72 layers (72 is chosen such that phases with period 2,3,4,6,8,9
etc will give a constant value).  To further reduce 
oscillation, the ratio is printed out only at layers with  $\sigma=+1$.
Substantial single phase regions can be seen, together with shorter 
transitional regions.  The overall trend towards a limiting value can 
be seen.

Each phase is stable only over a finite number of layers. Since the 
range of stability is inversely proportional to the period of the structure 
\cite{bak} the thickness over which some long-period structures are stable 
will be shorter than the periodic repeat distance of these structures.  
Consequently, they cannot be unambiguously identified. 

The long term trend of Figure 2 
is toward $<V>_m=-0.5$.  This corresponds to equal numbers of   
$\sigma=\pm 1$
which give a mean field value of $V=0$ averaged over all layers, and  
$<V>_m=-A/2=-0.5$ averaged over the   $\sigma= +1$ layers only.
A very curious phenomenon observed in Figure 2 
is that the ordered phases show 
{\it antiscreening } behavior: the mean value of $<V>_m$ moves away 
from the asymptotic value for most of range of the ordered phase.
Thus while the overall trend is an asymptotic increase of $<V>_m$, 
within any given phase $d<V>_m/dN$ is negative\cite{weird}.  
Evolution towards the asymptote 
is achieved via the boundaries between the phases, rather than the screening
by the phases themselves. 

\newpage
\begin{figure}[ht]
\leavevmode 
\epsfxsize=90mm
\begin{center}
\epsffile{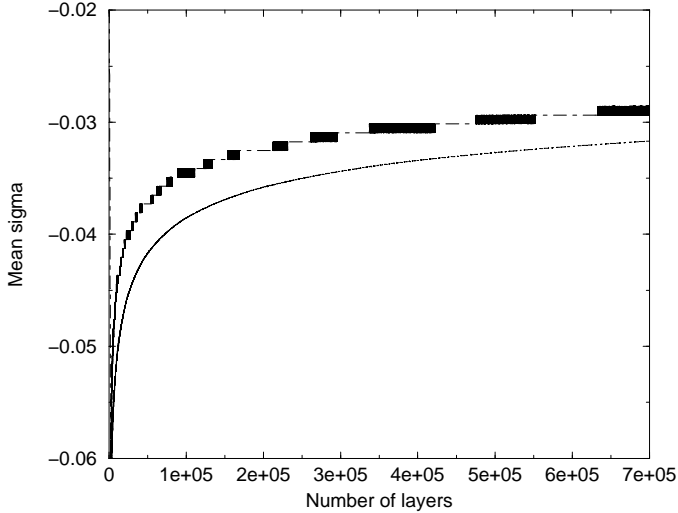}
\end{center}
\caption{Mean value of $\sigma$ for $\nu=1$, $A=1$.  The upper line is
averaged over the preceding 2520 layers, picking out as straight lines
phases of repeat period as in figure 1.  Transitions between these
short period phases are characterized by longer period phases, which
are shown by thick black lines (actually representing a
rapid fluctuation between K/2520 and (K+1)/2520 for integer K).  
The lower line is averaged over the whole layer, and shows the very 
slow monotonic growth of $<\sigma>$ towards the asymptotic value of 
$<\sigma>=0$).  The increase in the mean value of $1/<\sigma>$ is 
logarithmic, a reasonable fit to the graph being 
$<\sigma>= [7-2.9\ln N]^{-1}$.  For higher values of $A$ the convergence 
of $<\sigma> \rightarrow 0$ is even slower. For $\nu>1$ the asymptotic 
value of 
$<\sigma>$ is non-zero: it is a phase from the devil's staircase 
dependent upon $A$.	
\protect\label{fig:spin}}
\end{figure}

In mapping onto a real growth process, the value of $A$ is 
 determined by the material being deposited, however it may be possible 
by choice of substrate or external applied field to control the 
initial value of 
the potential.  By doing so, the density profile of $\sigma=+1$ may be varied.
This is not straightforward however: if the initial condition is compatible 
with the equilibrium structure a perfect multilayer can be grown, if not 
the concentration will traverse all possible phases with  $\sigma=+1$ 
density intermediate between the starting and equilibrium ones. 
There are an infinity of these phases, but their thickness 
must take an integer value, thus not all phases can actually be observed. 
If the interlayer spacing is 
taken to be of atomic dimension, a perfectly grown
(i.e. zero-T, zero entropy) film of even a millimeter thickness 
may not reach equilibrium and  
will appear disordered to any experimental probe.  

By interpreting the layer number $N$ as a time rather than a thickness, 
this type of growth-kinetic model also provides a simple model of 
history-dependence. In many social phenomena, decisions are 
made for one of two courses of action based on the evidence of 
past behavior with more recent evidence having a stronger 
weight\cite{voter}.  
Here the model already contains enough complexity to behave 
counterintuitively: the long term trend of increasing $<V>_m$ is opposite 
in sign to the $d<V>_m/dN$ measured over the stable phases, despite 
the fact that formally the devil's staircase provides a stable phase
at all $N$, and by implication $d<V>_m/dN$ is negative everywhere.
Of course, the spin-Ising model is a gross oversimplification of any 
real decision making process: this only serves to emphasize the 
non-trivial relation between the model and its behavior, and the 
difficulty for measurement when $d<V>_m/dN$ is negative everywhere
while $\Delta V / \Delta N$ is positive.

In sum, the growth kinetics of a simple  model exhibiting 
long range antiferromagnetic and short range ferromagnetic ordering
have been studied.  This has been shown to exhibit logarithmically slow 
equilibration to the equilibrium structure, 
passing through a formally infinite number of intermediate phases
which form a devil's staircase. 

The author would like to thank D.Srolovitz and J.Rickman for 
helpful discussions and hospitality at Princeton University, 
and support from the Fulbright Foundation.



\end{multicols}

\begin{references}
\bibitem{weird} The continuous 
limiting case $<V>_1$ is a function whose derivative
is negative everywhere, yet which increases to an asymptote by virtue 
of discontinuities at the (infinity of) phase boundaries.
\bibitem{voter} One can imagine the dilemma of a voter in the United
States, whose personal preference for (say) the Democrats ($A$) is
tempered by a belief that any party too long in government becomes
corrupt.  As with layer-growth, the inverse problem is extremely hard:  
study of the voting pattern is likely to conclude that it is random, 
rather than based on a simple decision making process.
\end{references}
\end{document}